\documentclass[twocolumn,aps,showpacs,preprintnumbers,amsmath,amssymb,floatfix]{revtex4}

\usepackage{graphicx}
\begin{document}

\title{Stretching of Proteins in the Entropic Limit}
\author{Marek Cieplak$^{1,2}$, Trinh Xuan Hoang$^3$, and
Mark O. Robbins$^1$}

\affiliation{
$^1$ Department of Physics and Astronomy, The Johns Hopkins University,
Baltimore, MD 21218\\
$^2$ Institute of Physics, Polish Academy of Sciences,
Al. Lotnik\'ow 32/46, 02-668 Warsaw, Poland \\
$^3$ The Abdus Salam International Center for Theoretical Physics,
Strada Costiera 11, 34014 Trieste, Italy}
\date{\today}

\begin{abstract}
{
Mechanical stretching of six proteins is studied through molecular dynamics
simulations. The model is Go-like, with Lennard-Jones interactions at native
contacts. Low temperature unfolding scenarios are remarkably complex and
sensitive to small structural changes. Thermal fluctuations reduce the peak
forces and the number of metastable states during unfolding. The unfolding
pathways also simplify as temperature rises.  In the entropic limit, all
proteins show a monotonic decrease of the extension where bonds rupture with
their separation along the backbone (contact order).
}
\end{abstract}

\maketitle

There is considerable current interest in the mechanical manipulation 
of single biological molecules. In particular, stretching studies
of large proteins with atomic force microscopes (AFM) and
optical tweezers \cite{Rief,Yang}
reveal intricate, specific, and reproducible
force ($F$) -- displacement ($d$) curves that call for
understanding and theoretical interpretation. 
The patterns depend on the pulling speed and on the stiffness of
the pulling device \cite{first}.
They must also depend on the effective temperature given by
the ratio of thermal to binding energies.
In experiments this ratio can be 
varied slightly by changing temperature $T$, and
over a large range by changing solvent properties\cite{Yang}, such as pH.
While the role of effective temperature in folding is well
studied \cite{socci,biophysical},
its effect on mechanical unfolding is not.

In this paper, we report results of molecular dynamics simulations 
of simplified models of six proteins that reveal universal trends
with increases in the effective temperature.
Thermal fluctuations accelerate rupture, lower the peaks in the 
F -- d curves, and reduce the number of peaks.
Most interestingly, the succession of unfolding events simplifies
from a complex pattern determined by the energy landscape
into a simple, uniform pattern determined by entropic considerations.
All of these changes are gradual, and move to completion near the 
$T$ where the specific heat peaks.
In the entropic limit,
rupturing events are governed exclusively
by the contact order, i.e. by the
distance along a sequence between
two amino acids which make a contact in the native state.
The contact order is also believed to be the major influence in 
folding to the native state.
However, there is no general correlation between
folding and the extremely complex unfolding scenarios observed
at low $T$ \cite{first}.

The models we use are coarse-grained and
Go-like \cite{Goabe}.
Full details are provided in earlier studies of folding
\cite{biophysical,Hoang}.
Briefly, the amino acids
are represented by point particles of mass $m$ located at the
positions of the C$^{\alpha}$ atoms.
They are tethered by a strong 
harmonic potential with a minimum at the peptide bond length.
The native structure of a protein is taken from the
PDB \cite{PDB} data bank and the interactions between the
amino acids are divided into native and non-native contacts.
The distinction is made by taking the fully atomic representation
of the amino acids in the native state and then associating native
contacts with overlapping amino acids.
The criterion for overlaps uses the
van der Waals radii of the atoms multiplied by 1.24
to account for the softness of the potential \cite{Tsai}.

The interaction between each pair of overlapping acids $i$ and $j$
is described with a 6-12 Lennard-Jones (LJ) potential
whose interaction length
$\sigma_{ij}$ (4.4 -- 12.8 \AA) is chosen so that the potential
energy minimum coincides with the native
C$^{\alpha}$ -- C$^{\alpha}$ distance.
This forces the ground state to coincide with the native state
at room $T$.
The non-native contacts are described by a LJ potential with
$\sigma =5$\AA $\;$ that is truncated at $r > 2^{1/6}\sigma$ to
produce a purely repulsive force.
An energy penalty is added to states with the wrong
chirality as described in Ref. \cite{biophysical}.
This facilitates folding, but has little effect on mechanical
stretching since the protein starts with the correct chirality.
All of the potentials have a common energy scale $\epsilon$,
which is taken as the unit of energy.
Time is measured in terms of the usual LJ vibrational time scale
$\tau \equiv \sqrt{m \sigma^2/\epsilon}$.

The desired effective temperature $\tilde T \equiv k_B T /\epsilon$,
where $k_B$ is Boltzmann's constant,
is maintained by coupling each C$^\alpha$ to a Langevin
noise \cite{Grest} and damping constant $\gamma$. The value of
$\gamma = 2 m/\tau$ is large enough to produce the overdamped
dynamics appropriate for proteins in a solvent \cite{biophysical},
but about 25 times smaller than the realistic damping 
from water. Previous studies show that this speeds
the dynamics by about a factor 25 without
altering behavior, and tests with larger $\gamma$ confirmed that
it merely rescales the diffusion time in the overdamped
regime \cite{Hoang}. As a result the effective value of $\tau$
in simulations is about 75 ps.

Stretching is accomplished by attaching both ends of the
protein to harmonic springs of spring constant
$k=0.12\epsilon/$\AA$^2$ \cite{first}.
This corresponds to a cantilever stiffness $k/2 \sim $0.2N/m, which
is typical of an AFM.
Stretching is implemented parallel to the initial
end-to-end position vector of the protein.
The outer end of one spring is held fixed at the origin,
and the outer end of the other is pulled at constant speed
$v_p = 0.005$ \AA$/\tau$.
Previous studies at $\tilde{T} = 0$ showed that decreasing
the velocity below this value had little effect on unfolding
\cite{first}.
This velocity corresponds to about $7\times 10^6$nm/s.
Velocities in all atom simulations are more than three orders of magnitude
faster\cite{Schulten},
but experimental AFM velocities are 1000 times slower.
Closing this gap remains a formidable challenge.
Varying the simulation temperature is one way to accelerate
experimental dynamics into an accessible range.

The displacement of the pulled end of the spring is denoted by $d$.
The net force stretching the protein is denoted by $F$,
and measured from the extension of the pulling spring.
Except at $\tilde{T}=0$, where the results are strictly reproducible,
$F$ is averaged over a displacement of 0.5\AA$\;$ 
to reduce thermal noise without substantially affecting spatial
resolution.
A contact between amino acids $i$ and $j$ is considered
ruptured if the distance between them exceeds 1.5 $\sigma _{ij}$.
The unfolding scenario is specified by the unbinding or
breaking distance $d_u$ for each native contact.
Note that at finite $\tilde{T}$ the contact may break and reform
several times and $d_u$ is associated with the final rupture.

\begin{figure}
\includegraphics[width=8cm]{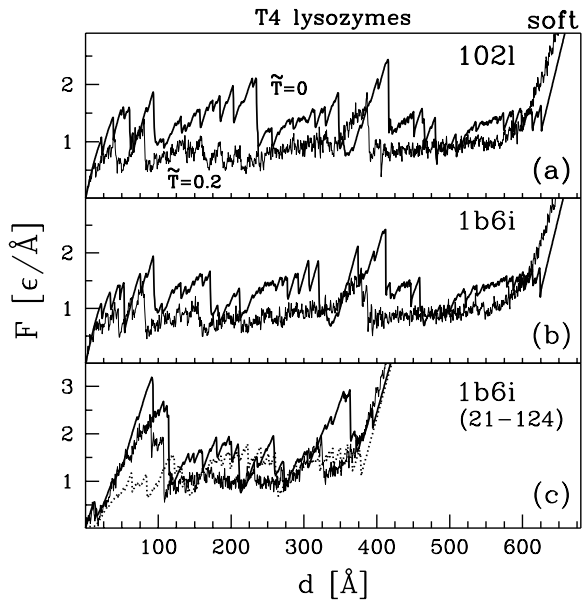}
\caption{
$F$--$d$ curves for three lysozyme systems: (a) 102l, (b) 1b6i
pulled from its ends and (c) 1b6i pulled by the cysteins at locations
21 and 124.
Thick and thin solid lines correspond to $\tilde{T}$=0 
and 0.2, respectively.
The dashed line in (c) shows $\tilde T= 0$
results for 1b6i with the 1-20 and 125-163 amino acids removed.
\label{fig:lysos}
 }
\end{figure}

\begin{figure}
\includegraphics[width=7.5cm]{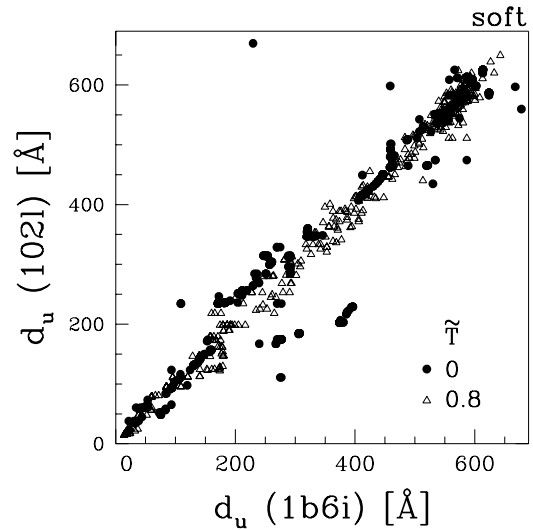}
\caption{
Breaking distances for bonds in 102l plotted against corresponding
values for 1b6i at the indicated $\tilde{T}$.
\label{fig:compare}
}
\end{figure}

To illustrate the sensitivity of low $\tilde{T}$ unfolding
to small structural changes, we consider bacteriophage T4 lysozymes.
There are two mutant structures with sequence length $N=$163
whose PDB codes are 1021 and 1b6i.
The latter has cysteins in locations 21 and 124 instead of threonine
and lysine respectively. 
Although the root mean square deviation between
the two structures is merely 0.3 \AA,
there are noticeable differences in the $\tilde{T}$=0
simulations of stretching in Fig. \ref{fig:lysos} (thick lines).
The force curves show a series of upward ramps where the protein
is stuck in a metastable state \cite{first}, followed by sharp drops as one
or more contacts break, allowing the intervening segment
to stretch. The differences between results for the two lysozymes,
especially around $d$ of 200, 375, and 475 \AA ,
indicate different sets of broken contacts.
Fig. \ref{fig:compare} compares the rupturing distances
of each native contact in the two mutants.
If the mutants followed the same pattern,
the points would lie on a line with unit slope.
However, they clearly follow different
patterns at $\tilde{T}=0$.
Sensitivity of $F-d$ curves to point mutations has been
demonstrated in recent experiments on an immunoglobulin
module in human cardiac titin \cite{Carrion}.

\begin{figure}
\includegraphics[width=7.5cm]{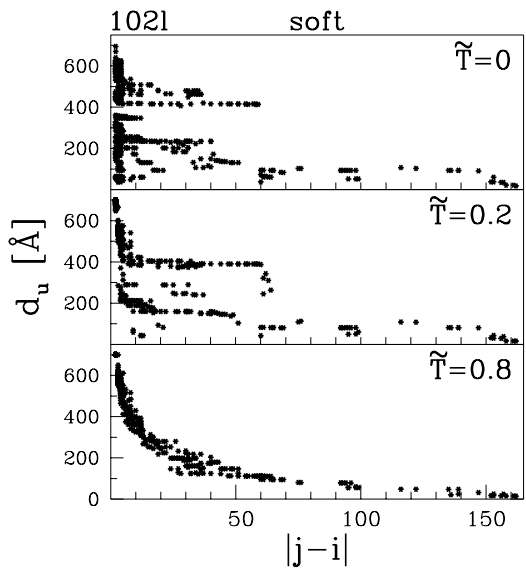}
\caption{
Contact breaking distances for 102l vs. contact order at the
indicated $\tilde{T}$.
\label{fig:contact}
}
\end{figure}
 
Figure \ref{fig:lysos} (c)
shows the importance of the location of the pulling force.
Here the pulling springs were attached to the cysteins at
$i$=21 and 124 of 1b6i, shortening the effective sequence length to
104 amino acids.
Yang et al. \cite{Yang} have used an AFM to study a string of 1b6i
proteins bound covalently at these sites, and observe a
series of equally spaced peaks that indicates the repeated units
unfold sequentially.
As shown in our earlier work \cite{first},
sequential unfolding occurs when the largest force peak breaks
the first contacts in a repeat unit.
Pulling from the $i$=21 and 124 sites produces a strong peak
near the start of unfolding, which is consistent with the
sequential unfolding observed by Yang et al..
This peak is absent for the full lysozymes and for the
sequence from $i$=21 to 124 with
amino acids $1 - 20$ and $125-163$ removed (dashed line).
We would thus predict repeated arrays of the full or
truncated sequences would unfold simultaneously if
linked at their ends.
It would be interesting to test these predictions with experiments.
Note that recent experiments on another protein, E2lip3,
have demonstrated that $F-d$ patterns can depend strongly on
the pulling geometry, particularly the direction of the force
relative to key native contacts \cite{Radford}.

We now turn to the effect of temperature on the force and unfolding
sequence.
Increasing $\tilde{T}$ to $0.2$
(thin lines in Fig. \ref{fig:lysos})
produces similar changes in the force curves for all lysozymes.
Thermal activation reduces the force needed to rupture bonds,
shifting the entire force curve downwards.
Some peaks disappear, indicating that the states are no longer metastable
at this stress and $\tilde{T}$.
Studies of the largest peaks show that they decrease roughly linearly
with $\tilde{T}$, and shift to smaller $d_u$.
For $\tilde{T} > 0.6$ no maxima can be identified and the
curves approach the entropic limit of a worm-like-chain (WLC) \cite{Doi}
at higher $\tilde{T}$.

\begin{figure}
\includegraphics[width=7.5cm]{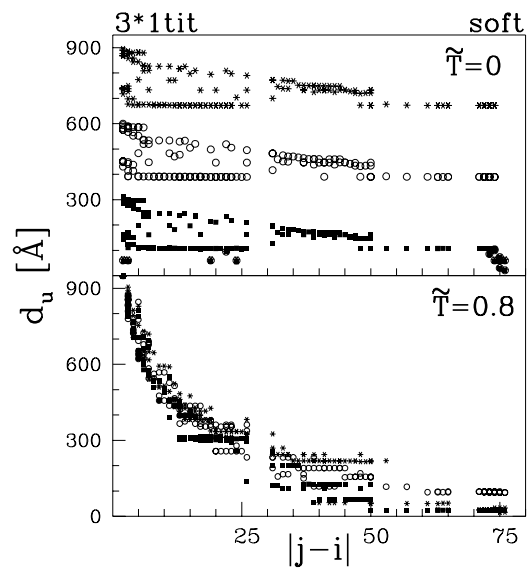}
\caption{
Unravelling of three serial domains of titin
on the
$d_u - |j-i|$ plane. Squares (stars) correspond to the most forward
(backward) domain.
\label{fig:titin3}
}
\end{figure}
\begin{figure}
\includegraphics[width=7.5cm]{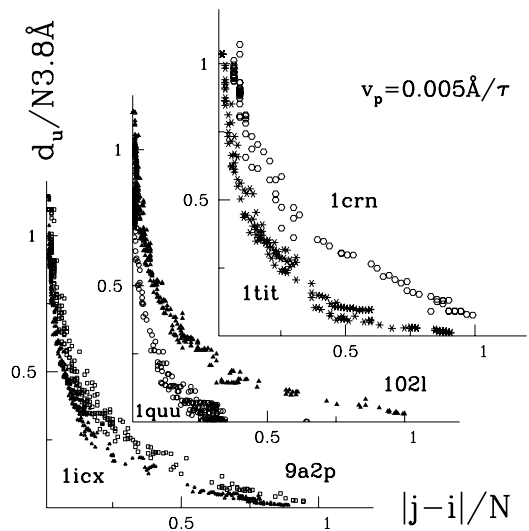}
\caption{
Unfolding in the high $\tilde{T}$ limit for
1crn (crambin; $N$=46), 1tit (the I27 domain of titin; $N=$89),
1quu (actinin; $N=$248), 102l (lysozyme; $N=$163), 
9a2p (barnase; $N=$108), and 1icx (yellow lupin protein 10; $N=$155). 
Both axes scaled by $N$ and $d_u$ is divided by the peptide
bond length of 3.8\AA to make it dimensionless.
For each protein the value of $\tilde{T}$ needed 
to reach the entropic limit
was near $\tilde{T}_{\rm max}$.  These values
were 0.6, 0.8, 1.2, 1.4, 0.8, and 1.2 respectively.
\label{fig:univ}
}
\end{figure}

The force curves and unfolding scenarios (Fig. \ref{fig:compare}) 
of the two mutant structures become more similar with
increasing $\tilde{T}$.
Similar plots of the unfolding sequences for
stiff vs. soft pulling springs also show considerable differences
at low $\tilde{T}$ that disappear as $\tilde{T}$ rises.
Fig. \ref{fig:contact} shows that the sequence of unfolding distances
also simplifies dramatically with increasing $\tilde{T}$.
Values of $d_u$ for each bond in 102l are plotted against
the bond's contact order $|j-i|$.
The $\tilde{T}=0$ unravelling scenario is complex:
Low contact order bonds break over the entire range of $d_u$,
while middle and long-range bonds break in
clusters of events at a few $d_u$.
At $\tilde{T} =0.2$, the events collapse onto a smaller
set of lines.
By $\tilde{T} =0.8$, events have nearly collapsed onto
a monotonically decreasing curve, that sharpens with further
increases in $\tilde{T}$.

The simplification in unfolding sequence is even more
dramatic for the tandem arrangement of three I27 domains
of titin shown in Fig. \ref{fig:titin3}.
At $\tilde{T}$=0, there is a serial unwinding
of the individual domains that produces a repeated pattern.
One domain unfolds from $d_u =$ 0 to 300\AA, the next from 300
to 600\AA, and the last from 600 to 900\AA.
The three sequences can be overlapped by a vertical displacement.
By $\tilde{T} =0.8$, unfolding occurs simultaneously and
$d_u$ has collapsed onto a nearly monotonic curve similar to
that found in Fig. \ref{fig:contact}.

We have examined the
$\tilde{T}$-dependence of unfolding
for a large number of proteins, including periodically repeated domains.
In all cases, increasing $\tilde{T}$ produces
a gradual reduction in the size and number of
force peaks and an increasingly universal relation between
$d_u$ and contact order. These changes
saturate near the temperature $\tilde{T}_{\rm max}$ where each
protein has a maximum in the equilibrium specific heat.
Fig. \ref{fig:univ} compares the normalized unfolding curves for
six proteins at temperatures near their $\tilde{T}_{\max}$.
For each protein, $d_u$ decreases nearly monotonically
at this $\tilde{T}$.
The curves are qualitatively similar, but results for
shorter proteins tend to lie above those for longer proteins.
Studies of the rate dependence show that this
is because longer proteins
take more time to sample configurations and thus are less
likely to reform contacts with large $|j-i|/N$.
The curves can be made more universal by choosing
different pulling rates for each protein.
Studies with artificial proteins (homopolymers), where every contact is
native, indicate a logarithmic rate dependence with the normalized
curves moving gradually up and to the right.

It is not surprising that $\tilde{T}_{\rm max}$ is the temperature
where unfolding simplifies.
It is where the entropy
of the system is changing most rapidly, and thus where
binding energies are becoming less important.
Not surprisingly,
the thermal energy at $\tilde{T}_{\rm max}$ is always comparable to the
contact energy $\epsilon$.
Other characteristic temperatures for folding are
substantially lower, and do not correlate well with
the simplification of the unfolding sequence.
For example, the temperature where folding is fastest,
$\tilde{T}_{\rm min}$, is 0.35 for our model of 102l and
the temperature where the protein spends half of its
time in the unfolded state is 0.25.
At both of these $\tilde{T}$'s the contact dependence
of the rupture process is still structured and non-monotonic,
as illustrated in Fig. \ref{fig:contact}.

In summary, thermal fluctuations affect the force -- displacement
curves in a profound manner, reducing force peaks and the number of
metastable configurations.
Yang et al. \cite{Yang} have observed the decrease in unbinding
force with increasing
$\tilde{T}$ by modifying
the solvent to reduce $\epsilon$.
However, as in our simulations for their tandem lysozyme system,
domains unfold serially and there is little structure except for
the initial force peak.
Studies of systems that unfold serially would exhibit richer
changes with $\tilde{T}$
and we predict this should be observed
for lysozymes joined at their ends.

Raising $\tilde{T}$
produces a dramatic simplification of the unfolding
sequence that culminates in a monotonic drop of $d_u$ with contact
order.
In the entropic limit, $\tilde{T} > \tilde{T}_{\rm max}$,
the tension in the chain is consistent with the
WLC model \cite{Doi}.
However, this model is often used to describe
unfolding at temperatures where folding is rapid.
In this regime we find substantial deviations from
the WLC model, which may 
affect the interpretation
of experimental data.
It would be desirable to repeat our studies with detailed atomistic
potentials like those used for titin \cite{Schulten}, but such studies
remain too computationally intensive for a thorough scan of parameter space.

This work was supported by by the Polish Ministry of Science
(grant 2 P03B 032 25)
and NSF Grant DMR-0083286.
MC appreciates discussions with A. Sienkiewicz.



\begin{thebibliography}{99}

\bibitem{Rief}
M. Rief, M. Gautel, F. Oesterhelt, J. M. Fernandez, and H. E. Gaub,
Science {\bf 276}, 1109-1112 (1997);
A. F. Oberhauser, P. E. Marszalek, H. P. Erickson, and J. M.  Fernandez, 
Nature {\bf 393}, 181-185 (1998);
M. Rief, J. Pascual, M. Saraste, and H. E. Gaub,
J. Mol. Biol. {\bf 286}, 553-561 (1999);
M. S. Z. Kellermayer, S. B. Smith, H. L. Granzier, C. Bustamante,
Science {\bf 276}, 1112-1116 (1997);
L. Tskhovrebova, K. Trinick, J. A. Sleep, M. Simmons,
Nature {\bf 387}, 308-312 (1997);
B. L. Smith, T. E. Schaeffer, M. Viani, J. B. Thompson, N. A. Frederic, 
J. Kindt, A. Belcher, G. D. Stucky, D. E. Morse, and P. K. Hansma,
Nature {\bf 399}, 761-763 (1999).


\bibitem{Yang}
G. Yang, C. Cecconi, W. A. Baase, I. R. Vetter, W. A. Breyer, J. A. Haack, 
B. W. Matthews, F. W. Dahlquist, and C. Bustamante,
Proc. Nat. Acad. Sci. {\bf 97}, 139-144 (2000).

\bibitem{first}
M. Cieplak, T. X. Hoang, and M. O. Robbins,
Prot. Struct. Func. Gen. {\bf 49}, 104-113 (2002);
114-124 (2002).

\bibitem{socci}
N. D. Socci and J. N. Onuchic, J. Chem. Phys. {\bf 101}, 1519-1528 (1994)

\bibitem{biophysical}
M. Cieplak and T. X. Hoang. 
Biophys. J. {\bf 84}, 475-488 (2003).

\bibitem{Goabe}
H. Abe, N. Go, 
Biopolymers {\bf 20}, 1013-1031 (1981);
S. Takada,
Proc. Natl. Acad. Sci. USA {\bf 96}, 11698-11700 (1999);
N. V. Dokholyan, S. V. Buldyrev, H. E. Stanley, and E. I. Shakhnovich,
Folding Des. {\bf 3}, 577-587 (1998).

\bibitem{Hoang}
T. X. Hoang and M. Cieplak,
J. Chem. Phys. {\bf 112}, 6851-6862 (2000);
{\bf 113}, 8319-8328 (2001).

\bibitem{PDB}
F. C. Bernstein, T. F. Koetzle, G. J. B. Williams,
E. F. Meyer Jr., M. D. Brice, J. R. Rodgers, O. Kennard, T. Shimanouchi,
and M. Tasumi,
J. Mol. Biol. {\bf 112}, 535-542 (1977).

\bibitem{Tsai}
J. Tsai, R. Taylor, C. Chothia, and M. Gerstein, J. Mol. Biol.
{\bf 290}, 253-266 (1999);
G. Settanni, T. X. Hoang, C. Micheletti, and A. Maritan,
Biophys. J. {\bf 83}, 3533-3541 (2002).

\bibitem{Grest}
G. S. Grest, K. Kremer, 
Phys. Rev. A {\bf 33}, 3628-3631 (1986).


\bibitem{Carrion}
H. Li, M. Carrion-Vazquez, A. F. Oberhauser,
P. E. Marszalek, and J. M. Fernandez,
Nature Struct. Biol. {\bf 7}, 1117-1120 (2000).

\bibitem{Radford}
D. J. Brockwell, E. Paci, R. C. Zinober, G. S. Beddard,
P. D. Olmsted, D. A. Smith, R. N. Perham, and S. E. Radford,
Nature Struct. Biol. {\bf 10}, 731 - 737 (2003).

\bibitem{Doi}
M. Doi, and S. F. Edwards, {\it The Theory of Polymer Dynamics},
Oxford U.P., Oxford, England (1989).

\bibitem{Schulten}
H. Lu, B. Isralewitz, A. Krammer, V. Vogel, K. Schulten,
Biophys. J. {\bf 75}, 662-671 (1998).




\end{thebibliography}
\end{document}